\documentclass[12pt]{iopart}

\usepackage[utf8]{inputenc}
\usepackage[T1]{fontenc}

\usepackage{iopams}
\usepackage{amsfonts}
\usepackage{amssymb}
\usepackage{graphicx}
\usepackage{hyperref}
\hypersetup{
	colorlinks=true,
	linkcolor=blue,
	filecolor=magenta,      
	urlcolor=cyan,
	pdftitle={Optimizing Active Seismic Isolation Systems in Gravitational-Wave Detectors},
	pdfpagemode=FullScreen,
}
\usepackage{subcaption}
\graphicspath{{figures}}
\usepackage{xcolor}

\begin{document}

	\title{Optimizing Active Seismic Isolation Systems in Gravitational-Wave Detectors}
	\author{
		Terrence Tsang$^{1,a}$,
		Fabi\'an E.~Pe\~na Arellano$^{2}$
		Takafumi Ushiba$^3$,
		Ryutaro Takahashi$^4$,
		Yoichi Aso$^4$,
		Katherine L. Dooley$^1$
	}
	\address{$^{1}$Cardiff University, Cardiff CF24 3AA, United Kindom}
\address{$^{2}$California State University, Los Angeles, College of Engineering, Computer Science and Technology, 5151 State University Dr, Los Angeles, CA 90032, USA.}
	\address{$^{4}$Gravitational Wave Science Project, National Astronomical Observatory of Japan, 2-21-1 Osawa, Mitaka City, Tokyo 181-8588, Japan}
	\ead{$^{a}$\href{mailto:TsangTT@cardiff.ac.uk}{TsangTT@cardiff.ac.uk}}
	
%
%
%
%
%

	\vspace{10pt}
	
	\begin{abstract}
		Gravitational wave detectors such as KAGRA, a 3-km long underground laser interferometer in Japan, require elaborate passive and active seismic isolation of their mirrors.
		With the aim of detecting passing gravitational waves that create a relative mirror displacement of less than $10^{-22}\ \mathrm{m}$ at frequencies of tens to hundreds of Hz, all environmental couplings must be stringently suppressed.
		This paper presents the result of applying the H-infinity optimization method to the active seismic isolation of a gravitational-wave detector for the first time.
		The so-called \textit{sensor correction} and \textit{sensor fusion} schemes of the seismic attenuation system of KAGRA's signal recycling mirror are used as a test bed.
		We designed and implemented optimal sensor correction and sensor fusion filters, resulting in a sevenfold attenuation of seismic noise coupling to the signal recycling mirror in the 0.1-0.5 Hz band, with the downstream effect of an 88.2\% noise performance improvement in the same frequency band.
		When combined with other hardware upgrades, the implementation of sensor correction and sensor fusion contributed to an increase in KAGRA's duty cycle from 53\% in the O3GK observation run to 80\% in O4a, demonstrating the effectiveness of the H-infinity optimization approach.
		
	\end{abstract}
	
	\section{Active Seismic Isolation in Gravitational-Wave Detectors}
\begin{figure}[!ht]
		\centering
		\includegraphics[width=.4\linewidth]{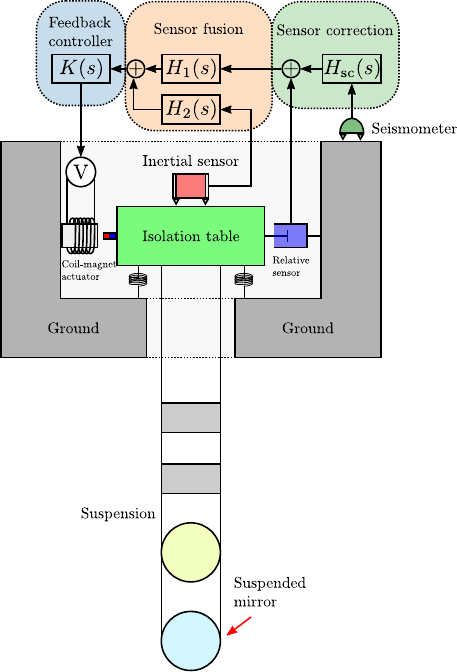}
		\caption{A conceptual illustration of an active isolation system suspending a mirror in a gravitational-wave detector.}
		\label{fig:active_isolation}
	\end{figure}

	Gravitational-wave detectors, such as Advanced LIGO \cite{advanced_ligo}, Advanced Virgo \cite{advanced_virgo}, and KAGRA \cite{kagra}, rely on ultra-high-precision laser interferometry to detect gravitational waves.
	If not sufficiently suppressed, low-frequency seismic disturbances, such as earthquakes and the secondary microseism caused by ocean storms, can cause misalignment of the interferometer's optical cavities, ultimately resulting in reduced sensitivity to gravitational waves and lock loss \cite{performance_of_kagra, Schwartz_2020}.
	To mitigate these effects, the main optics of the interferometer are isolated from the ground in a two-stage system: a multi-stage pendulum passively isolates the mirror, while an active isolated platform serves as the suspension point of for the pendulum \cite{ligo_seismic, kagra_type_bp, kagra_type_b}.
	A representative schematic of such a seismic isolation system is shown in Fig.~\ref{fig:active_isolation}.
	
	Active seismic isolation is provided through the use of \textit{feedback} and \textit{feedforward} control systems, which rely on an array of sensors and actuators.
	The sensors include relative displacement sensors that measure the differential motion between the isolation table and the ground; seismometers that measure the ground motion; and inertial sensors which provide a measurement of the table's inertial motion.
	Control filters turn these witnesses into feedback and feedforward signals which are sent to actuators on the tables, such as coil-magnet devices.
	
	
	The goal is to provide a table that is as inertial as possible, which equates to a need to create as a high-fidelity measurement of inertial motion as possible from the collection of sensors, each of which has its own distinct noise floor, frequency range of greatest sensitivity, and function.
	By combining the sensor signals, an inertial ``super sensor'' can be created.
	The combination takes two forms: \textit{sensor correction} and \textit{sensor fusion}.
	The sensor correction scheme combines information from the seismometer and the relative displacement sensors to create a new displacement signal, effectively free from seismic noise \cite{ligo_seismic}.
	These sensor-corrected relative signals are then combined with inertial sensor signals through sensor fusion \cite{ligo_seismic, myself, mythesis} to arrive at the inertial super sensor.
	It is the design of the filters used to create this super sensor that this paper focuses on.

	
	There are three design freedoms in an active isolation systesm: the feedback controller $K(s)$, complementary filters $H_1(s)$ and $H_2(s)$, and the sensor correction filter $H_\mathrm{sc}(s)$.
	These control filters are infinite impulse response (IIR) filters that determine the frequency-dependent performance of the active isolation system.
	Given that the instruments involved in the control process are not noiseless, designing these filters requires an optimal trade-off between seismic noise suppression and sensor noise suppression.
	
	Conventionally, these control filters are designed by manually shaping their frequency responses, often through an iterative process of placing zeros and poles.
	This labor-intensive task can result in inconsistent and suboptimal performance.
	To address these challenges, an optimization-based approach is proposed.
	
	The $\mathcal{H}_\infty$ (H-infinity) method has proven to be highly effective for optimizing complementary filters in sensor fusion configurations \cite{myself, mythesis, thomas}.
	The noise floor of an H-infinity super sensor approaches the instrumentation limit, exhibiting minimal deviation from the lower noise spectrum of the two sensors.
	In this work, we revisit the optimal sensor fusion method and extend its application to address the sensor correction and feedback control problems in seismic isolation systems for gravitational wave detection.
	
	In KAGRA, sensor correction and fusion were not implemented during O3GK due to the challenges associated with manually shaping appropriate control filters.
	As a result, the duty cycle of KAGRA was significantly impacted by secondary microseismic activities during O3GK \cite{performance_of_kagra}.
	The application of the H-infinity method led to the successful implementation of sensor correction and fusion in the 4 Type-B suspensions in KAGRA, which isolate the beam splitter optics, and the signal-recycling mirrors.
	This achievement contributed to an improvement in the duty cycle, from 53\% during O3GK \cite{performance_of_kagra} to 80\% in O4a \cite{status_of_underground}.

	The paper is organized as follows:
	Section~\ref{sec:method} reviews the H-infinity optimal sensor fusion method and extends the application of the H-infinity method to sensor correction and feedback control.
	Section~\ref{sec:results} presents the results of the H-infinity sensor correction and sensor fusion methods to KAGRA's seismic isolation systems.
	Section~\ref{sec:discussions} addresses the limitations and potential future enhancements of the H-infinity method.
	Finally, Section~\ref{sec:conclusions} provides the concluding remarks.
	
	\section{H-infinity Seismic Isolation\label{sec:method}}
	
	\subsection{Optimal Sensor Fusion\label{sec:optimal_sensor_fusion}}
	
	\begin{figure}[!ht]
		\centering
		\begin{subfigure}{.49\linewidth}
			\centering
			\includegraphics[width=1\linewidth]{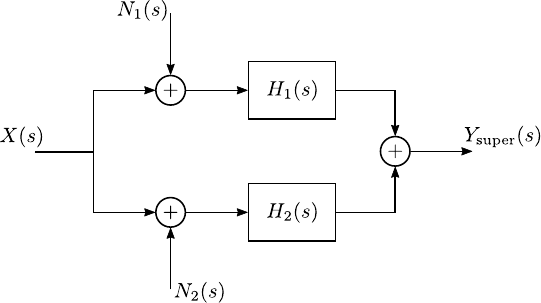}
			\caption{A sensor fusion configuration using complementary filters.}
			\label{fig:sensor_fusion}
		\end{subfigure}\hfill
		\begin{subfigure}{.49\linewidth}
		\centering
		\includegraphics[width=\linewidth]{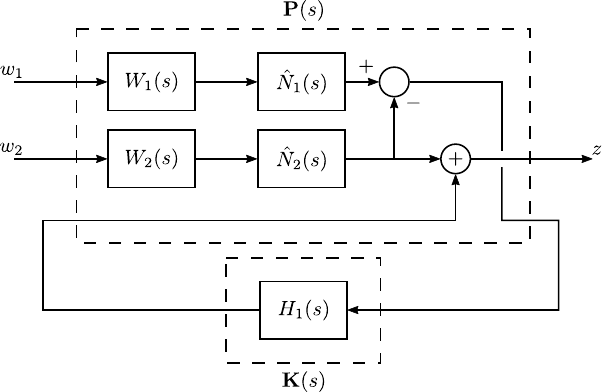}
		\caption{Generalized plant representation of the complementary filter problem.}
		\label{fig:generalized_plant}
		\end{subfigure}
		\caption{Sensor fusion block diagrams.}
	\end{figure}
	
	Fig.~\ref{fig:sensor_fusion} illustrates the sensor fusion configuration utilizing complementary filters.
	A common signal, $X(s)$ is measured by two sensors, each of which introduces sensor noise, $N_1(s)$ and $N_2(s)$, respectively.
	The two measurements are processed through complementary filters, $H_1(s)$ and $H_2(s)$.
	These filtered signals are then combined to produce a super sensor readout, $Y_\mathrm{super}(s)$, expressed as:
	\begin{equation}
		Y_\mathrm{super}(s) = \left[H_1(s)+H_2(s)\right]X(s) + N_\mathrm{super}(s)\,,
		\label{eqn:super_sensor}
	\end{equation}
	where the combined sensor noise, $N_\mathrm{super}(s)$ is given by:
	\begin{equation}
		N_\mathrm{super}(s) = H_1(s)N_1(s) + H_2(s)N_2(s)\,.
		\label{eqn:combined_sensor_noise}
	\end{equation}
	
	In a sensor fusion setting, the combined sensor noise is fundamentally limited by the individual noise spectra, $N_1(s)$ and $N_2(s)$.
	The corresponding instrumentation limit, i.e. best achievable noise floor, of the combined sensor noise spectrum is given by $\min(|N_1(j\omega)|, |N_2(j\omega)|)$.
	An optimal sensor fusion configuration employs complementary filters designed such that the spectrum $|N_\mathrm{super}(j\omega)|$ deviates minimally from this instrumentation limit.
	This deviation serves as an effective performance metric and can be utilized as a cost function for optimization.
	
	The deviation is a frequency-dependent quantity that can be characterized by the \textit{H-infinity norm}, which represents the supremum of the deviation across all frequencies.
	Through optimization, the deviation is minimized uniformly across the frequency spectrum, becoming frequency-independent.
	This indicates that the spectrum of the combined sensor noise closely follows the instrumentation limit.
	Such behavior is ideal for applications in gravitational-wave detectors, where frequency-dependent noise specifications can vary by orders of magnitude across the spectrum, making the H-infinity method an ideal choice.
	
	To use the H-infinity method, the sensor fusion configuration is expressed in a generalized plant representation, as illustrated in Fig.~\ref{fig:generalized_plant}.
	In this represenation, the frequency content of the noise sources $N_1(s)$ and $N_2(s)$ are modeled by transfer functions $\hat{N}_1(s)$ and $\hat{N}_2(s)$, respectively.
	The weighting functions $W_1(s)$ and $W_2(s)$ define frequency-dependent targets for the filtered noises.
	Here, $H_2(s)$ is replaced by $1-H_1(s)$, under the assumption $H_1(s)+H_2(s)=1$, ensuring no signal distortion as described in Eqn.~\eref{eqn:super_sensor}.
	
	The closed-loop transfer matrix of the generalized plant is expressed as
	\begin{equation}
		\mathbf{G}(s) =
		\left[
		\begin{array}{cc}
			H_1(s)\hat{N}_1(s)W_1(s) & \left[1-H_1(s)\right]\hat{N}_2(s)W_2(s)
		\end{array}
		\right]\,.
		\label{eqn:transfer_matrix}
	\end{equation}
	The H-infinity norm is defined as
	\begin{eqnarray}
		\left\Vert \mathbf{G}(s)\right\Vert_\infty &= \sup_\omega \sqrt{\lambda\left(\mathbf{G}(j\omega)\mathbf{G}(j\omega)^H\right)}\,,
	\end{eqnarray}
	where $\lambda(\cdot)$ denotes the eigenvalue.
	
	For frequencies where the contribution from first element of the transfer matrix in Eqn.~\eref{eqn:transfer_matrix} dominates over the second element, the H-infinity norm can be approximated as
	\begin{equation}
		\left\Vert \mathbf{G}(s)\right\Vert_\infty \approx \sup_\omega\left|H_1(j\omega)\hat{N}_1(j\omega)W_1(j\omega)\right|\,.
		\label{eqn:h_norm_approx}
	\end{equation}
	If optimization yields an H-infinity norm of value $\Vert \mathbf{G}(s)\Vert_\infty = \gamma$, Eqn.~\eref{eqn:h_norm_approx} can then be rewritten as
	\begin{equation}
		\left| H_1(j\omega)\hat{N}_1(j\omega)\right| \leq \gamma \left|W_1(j\omega)\right|^{-1}\,.
	\end{equation}
	This implies that $\gamma |W_1(j\omega)|^{-1}$ serves as the upper bound of the filtered sensor noise $H_1(s)N_1(s)$.
	Likewise, a similar inequality can be established at all frequencies for $H_2(s)N_2(s)$, leading to $\gamma |W_2(j\omega)|^{-1}$ as its upper bound.
	Since the inequality holds true across all frequencies, the reciprocals of $W_1(s)$ and $W_2(s)$ defines the frequency-dependent specifications for the filtered noises.
	
	For optimal sensor fusion, the combined sensor noise should deviate minimally from the instrumentation limit, $N_1(s)$ or $N_2(s)$, whichever is lower.
	A natural choice for the weighting functions is then\\
	\begin{minipage}{.5\linewidth}
	\begin{equation}
		W_1(s) = \frac{1}{\hat{N}_2(s)}\,,
	\end{equation}
	\end{minipage}
	\begin{minipage}{.5\linewidth}
		\begin{equation}
			W_2(s) = \frac{1}{\hat{N}_1(s)}\,.
		\end{equation}
	\end{minipage}
	This ensures that the weighting functions emphasize the less noisy sensor at each frequency, leading to an optimal fusion that closely follows the instrumentation limit.
	
	Under this setting, the H-infinity norm $\gamma$, which is minimized in the optimization process, quantifies the deviation from the instrumentation limit.
	This directly aligns with the goal established earlier in this section, achieving optimal sensor fusion by minimizing deviations from the lowest achievable noise level.
	
	In this work, our \verb|Kontrol| \cite{kontrol} Python package is used to synthesize an H-infinity optimal controller for the generalized plant $\mathbf{P}(s)$.
	The \verb|Kontrol| package is a wrapper around Python Control \cite{pythoncontrol}, which employs the \verb|SB10AD| Fortran subroutine, an implementation of the solutions in Refs.~\cite{statespace_solutions_h_infinity, Glover1988StatespaceFF, fortran77_continuous_time_h_infinity}.

	\subsection{Optimal Sensor Correction and Feedback Control}
	
	In principle, sensor fusion, sensor correction, and feedback control can each be addressed separately using distinct fomulations, such as the mixed-sensitivity synthesis method for feedback control \cite{multivariable_skogestad}.
	However, these three control problems share a similar mathematical structure, each involving minimizing two conflicting objective functions using complementary filters.
	Given this commonality, they can be unified under a single optimization framework, as described in Sec.~\ref{sec:optimal_sensor_fusion}, by appropriately substituting variables.
	
	To demonstrate the mathematical equivalence among the three problems, we will derive the objective functions for noise minimization in sensor correction and feedback control, expressing them in the same form as Eqn.~\eref{eqn:combined_sensor_noise}.
	
	\begin{figure}[!h]
		\centering
		\begin{subfigure}{.49\linewidth}
		\centering
		\includegraphics[width=\linewidth]{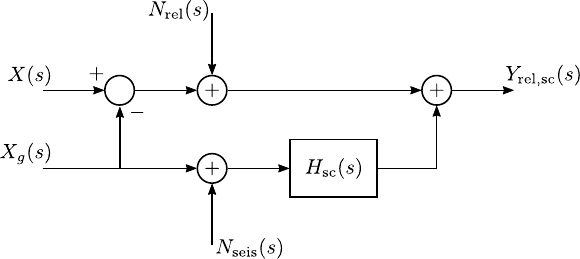}
		\caption{Sensor correction for relative displacement sensors.}
		\label{fig:sensorcorrection}
		\end{subfigure}
		\hfill
		\begin{subfigure}{.49\linewidth}
			\centering
			\includegraphics[width=\linewidth]{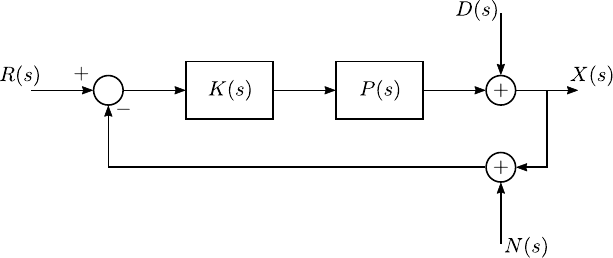}
			\caption{Feedback control with disturbance and sensing noise.}
			\label{fig:feedback_control}
		\end{subfigure}
		\caption{Sensor correction and feedback control block diagrams.}
	\end{figure}
	
	\subsubsection{Sensor correction}
	Fig.~\ref{fig:sensorcorrection} illustrates a sensor correction configuration for relative displacement sensors.
	In this setup, the differential displacement $X(s) - X_g(s)$ between the isolation table and the ground is measured by a relative displacement sensor, which introduces measurement noise $N_\mathrm{rel}(s)$.
	The undesired ground motion coupling, $X_g(s)$, can be mitigated using a seismometer measurement of the ground motion, which itself is subject to measurement noise $N_\mathrm{seis}(s)$.
	To prevent excess noise injection, the seismometer signal is processed through a sensor correction filter $H_\mathrm{sc}(s)$ before being combined with the relative sensor signal.
	
	The sensor-corrected relative sensor output is given by
	\begin{equation}
		Y_\mathrm{rel, sc}(s) = X(s) + N_\mathrm{rel,sc}(s)\,,
	\end{equation}
	where the sensor correction noise is
	\begin{equation}
		N_\mathrm{rel,sc}(s) = N_\mathrm{rel}(s) + H_\mathrm{sc}(s)N_\mathrm{seis}(s) - \left[1-H_\mathrm{sc}(s)\right]X_g(s)\,.
		\label{eqn:sensor_correction_noise}
	\end{equation}
	The objective is to minimize $N_\mathrm{rel,sc}(s)$, ensuring that the sensor correction effectively reduces seismic noise coupling while avoiding excess noise injection from the seismometer.
	
	In Eqn.~\eref{eqn:sensor_correction_noise}, the sensor correction filter $H_\mathrm{sc}(s)$ does not influence the relative sensor noise $N_\mathrm{rel}(s)$, which remains as an ambient term contributing to the instrumentation limit.  
	The remaining components of Eqn.~\eref{eqn:sensor_correction_noise} closely resemble Eqn.~\eref{eqn:combined_sensor_noise}, where two noise sources, $N_\mathrm{seis}(s)$ and $X_g(s)$, are suppressed by complementary terms, $H_\mathrm{sc}(s)$ and $1 - H_\mathrm{sc}(s)$, respectively.
	This structural similarity implies that the sensor correction problem can be formulated using the optimal sensor fusion method from Sec.~\ref{sec:optimal_sensor_fusion}, with the following substitutions:	$N_1(s) = N_\mathrm{seis}(s)$, $N_2(s) = X_g(s)$, and $H_1(s) = H_\mathrm{sc}(s)$.
	
	When the relative sensor noise is small compared to the seismometer noise and the ground motion, the instrumentation limit can be fully specified by  $W_1(s) = 1/\hat{X}_g(s)$ and $W_2(s) = 1/\hat{N}_\mathrm{seis}(s)$, where the hat notation denotes the transfer function models of the ground motion and the seismometer noise, respectively.  
	Alternatively, when the relative sensor noise is large compared to the seismometer noise and the ground motion, the instrumentation limit simplifies to  
	$W_1(s) = W_2(s) = 1/\hat{N}_\mathrm{rel}(s)$, where $\hat{N}_\mathrm{rel}(s)$ represents the relative sensor noise model.  
	More generally, when all noise sources are equally significant, the instrumentation limit follows $\max(|N_\mathrm{rel}(j\omega)|, \min(|N_\mathrm{seis}(j\omega)|, |X_g(j\omega)|))$.
	In this scenario, an explicit model of the instrumentation limit is required to appropriately define the weighting functions.\\
	
	\subsubsection{Feedback control}
	Fig.~\ref{fig:feedback_control} illustrates a feedback control configuration used in active seismic isolation systems for gravitational-wave detectors. In this framework, the displacement of the isolation table, $X(s)$, is given by  
	\begin{equation}
	X(s) = \frac{1}{1+K(s)P(s)}D(s) - \frac{K(s)P(s)}{1+K(s)P(s)}N(s)\,,
	\label{eqn:feedback_control}
	\end{equation}
	where $K(s)$ is the feedback controller, $P(s)$ is the plant that characterizes the dynamics of the isolation table, $D(s)$ represents seismic disturbances, and $N(s)$ is the measurement noise.
	The setpoint $R(s)$ is omitted since it is typically a DC term without significant frequency content.
	
	Like sensor correction, the isolation table displacement described in Eqn.~\eref{eqn:feedback_control} is analogous to Eqn.~\eref{eqn:combined_sensor_noise}.
	By substituting $N_1(s) = D(s)$, $N_2(s) = N(s)$, and $H_1(s) = S(s) \equiv 1/[1+K(s)P(s)]$, the H-infinity controller can be derived as  
	\begin{equation}
	K(s) = \frac{1}{P(s)}\frac{1-S(s)}{S(s)},
	\end{equation}
	using the same H-infinity formulation outlined in Sec.~\ref{sec:optimal_sensor_fusion}.
	 
	It is worth noting that the H-infinity optimal feedback control problem is more commonly formulated in a mixed-sensitivity setting, where the controller $K(s)$ is optimized directly rather than the sensitivity function $S(s)$ \cite{multivariable_skogestad}.
	One advantage of framing the feedback control problem as a sensor fusion problem is that it results in a lower-order generalized plant, improving numerical accuracy in the optimization.
	Nevertheless, both approaches are mathematically equivalent, ensuring that H-infinity optimization yields the same optimal controller.\\
	 
	\section{Optimizing KAGRA Seismic Attenuation Systems\label{sec:results}}
	
	\begin{figure}[!h]
		\centering
		\includegraphics[width=0.4\linewidth]{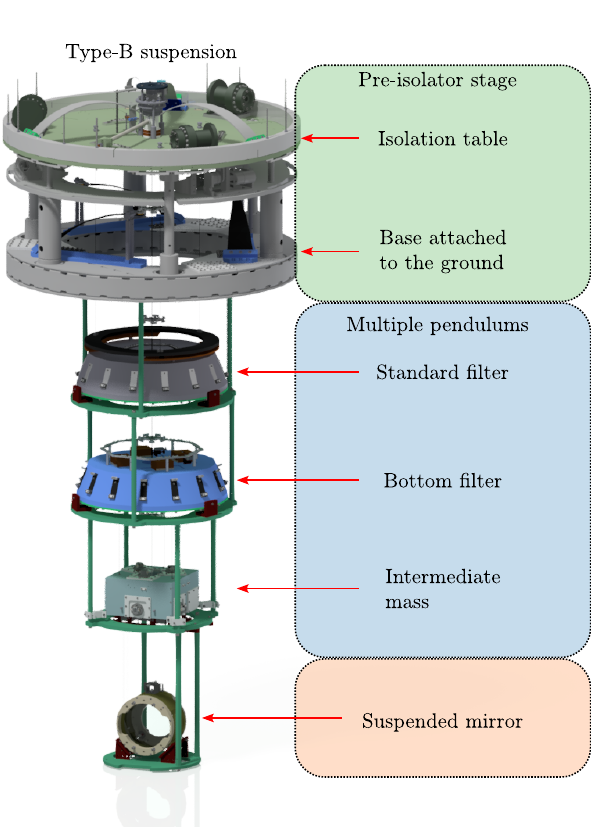}
		\caption{Type-B Seismic Attenuation Systems}
		\label{fig:typebsuspension}
	\end{figure}
	
	The Type-A (isolating the 4 test mass optics) and Type-B seismic attenuation systems (suspensions) in KAGRA utilize a pre-isolator stage with active seismic isolation capabilities.
	Fig.~\ref{fig:typebsuspension} illustrates a Type-B suspension, which is the type of suspension used to suspend the beamsplitter and the signal recycling mirrors in KAGRA.
	The pre-isolator stage is equipped with linear variable differential transformers (LVDTs) to measure the differential displacement between the isolation table and the ground.
	Additionally, L-4C geophones from Sercel are installed on the isolation table to measure inertial velocity, while a Trillium Compact seismometer is placed nearby to monitor ground motion.
	These three sensors are integrated into the sensor correction and sensor fusion configurations.
	
	During O3GK, sensor correction and sensor fusion configurations were not implemented due to the difficulty of designing sensor correction and complementary filters that met stringent noise constraints.
	As a result, KAGRA was significantly affected by secondary microseismic disturbances, limiting its duty cycle to 53\% \cite{performance_of_kagra}.  
	
	Following O3GK, several upgrades were made to the KAGRA detectors \cite{status_of_underground}, including the installation of folded pendulum accelerometers on the Type-A isolation table, and the implementation of sensor correction and sensor fusion on both Type-A and Type-B suspensions.
	These improvements contributed to a substantial increase in KAGRA’s duty cycle, reaching 80\% during O4a.  
	
	This section outlines the design of sensor correction and complementary filters for the Type-B suspensions using the H-infinity method.
	The two control schemes were implemented as shown in Fig.~\ref{fig:active_isolation}, serving as the readout for the active isolation feedback control.
	
	\subsection{H-infinity Sensor Correction}
	\begin{figure}[!h]
		\centering
	\begin{subfigure}{.49\linewidth}
		\centering
		\includegraphics[width=\linewidth]{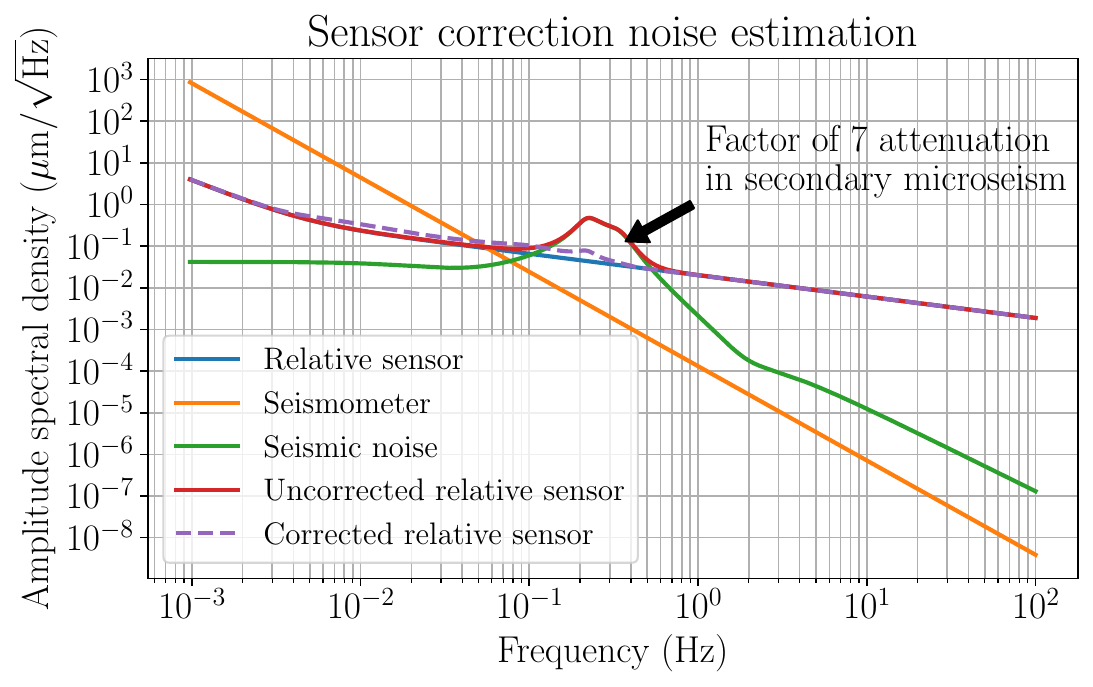}
		\caption{Sensor and seismic noises.}
		\label{fig:sensorcorrectionestimation}
	\end{subfigure}
	\begin{subfigure}{.49\linewidth}
		\centering
		\includegraphics[width=\linewidth]{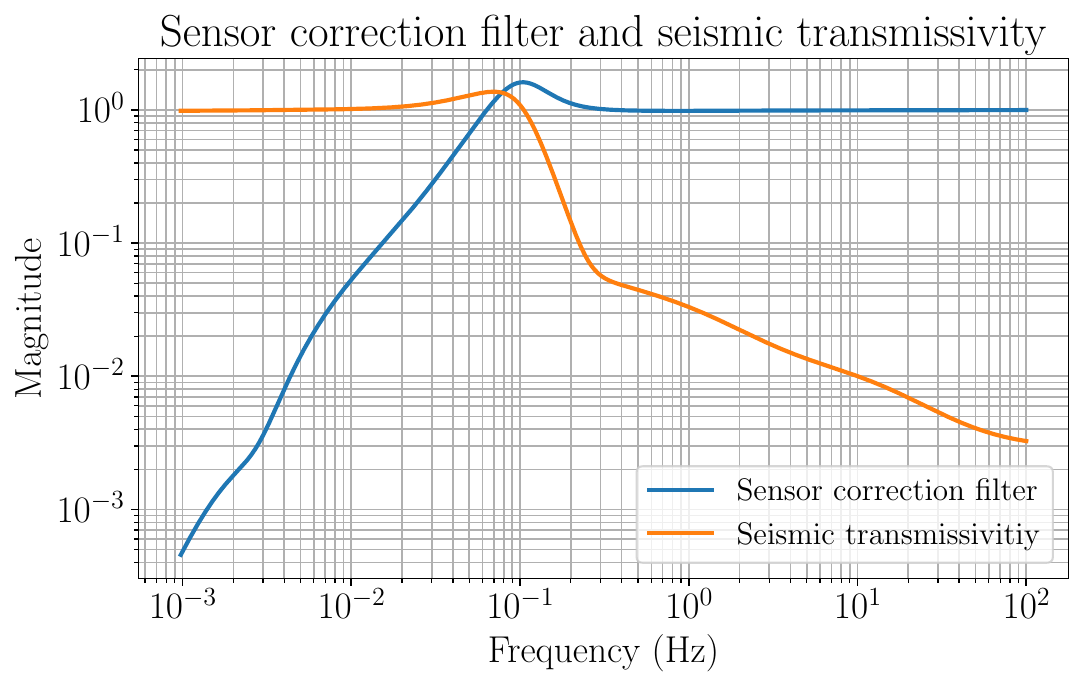}
		\caption{H-infinity optimal sensor correction fliter.}
		\label{fig:sensorcorrectionfilter}
	\end{subfigure}
	\caption{Noise spectrums and control filters in the sensor correction scheme.}
	\end{figure}

	Fig.~\ref{fig:sensorcorrectionestimation} illustrates the sensor and seismic noises involved in the sensor correction configuration.
	In this case, the relative sensor noise is higher than both the seismometer noise and the seismic noise.
	Therefore, the relative sensor noise model serves as the frequency-dependent target in the weighting functions.
	
	The H-infinity optimal sensor correction filter and its complement (transmissivity, i.e. $1-H_\mathrm{sc}(s)$) are shown in Fig.~\ref{fig:sensorcorrectionfilter}.
	As observed, the H-infinity method produces a high-pass sensor correction filter, consistent with traditional designs.
	The filter crosses unity at approximately 0.06 Hz, aligning with the frequency at which the seismometer and relative sensor noise levels intersect.
	The transmissivity behaves as a low-pass filter, with a unity crossing around 0.1 Hz, where seismic noise contributions begin to dominate relative sensor noise.
	It provides approximately a tenfold suppression of seismic noise at 0.2 Hz, coinciding with the secondary microseismic peak.
	
	The optimized filters are used to estimate the sensor noise of the sensor-corrected relative sensor, shown as a dashed line in Fig.~\ref{fig:sensorcorrectionestimation}.
	As observed, seismic noise coupling in the relative readout is significantly suppressed, with up to a sevenfold reduction at the microseismic peak.
	The sensor correction scheme introduces only a minimal amount of seismometer noise below the microseismic band.
	
	As observed, the sensor noise of the corrected sensor deviates minimally from the original relative sensor noise, consistent with the H-infinity norm of the system, $\gamma = 1.033$.
	Since the relative sensor noise serves as the frequency-dependent target, this indicates that the sensor-corrected relative sensor has a measurement noise level at most 1.033 times that of the original relative sensor noise.
	
	The relative sensor without sensor correction (uncorrected relative sensor) measures the relative displacement between the isolation platform and the ground.
	With the dispalcement of the isolation platform being the variable of interest, the uncorrected relative sensor noise is defined as the superposition of the intrinsic relative sensor noise and the seismic noise, i.e. what the noise floor would have been without sensor correction.
	The root mean squared (RMS) value of the uncorrected relative sensor noise is 0.192 $\mu\mathrm{m}$, while that of the sensor-corrected relative sensor is 0.120 $\mu\mathrm{m}$, representing a 37.5\% reduction.
	This demonstrates the efficacy of the H-infinity method in suppressing seismic noise.
	
	\subsection{H-infinity Sensor Fusion}

	\begin{figure}[!h]
		\centering
		\begin{subfigure}{.49\linewidth}
			\centering
			\includegraphics[width=\linewidth]{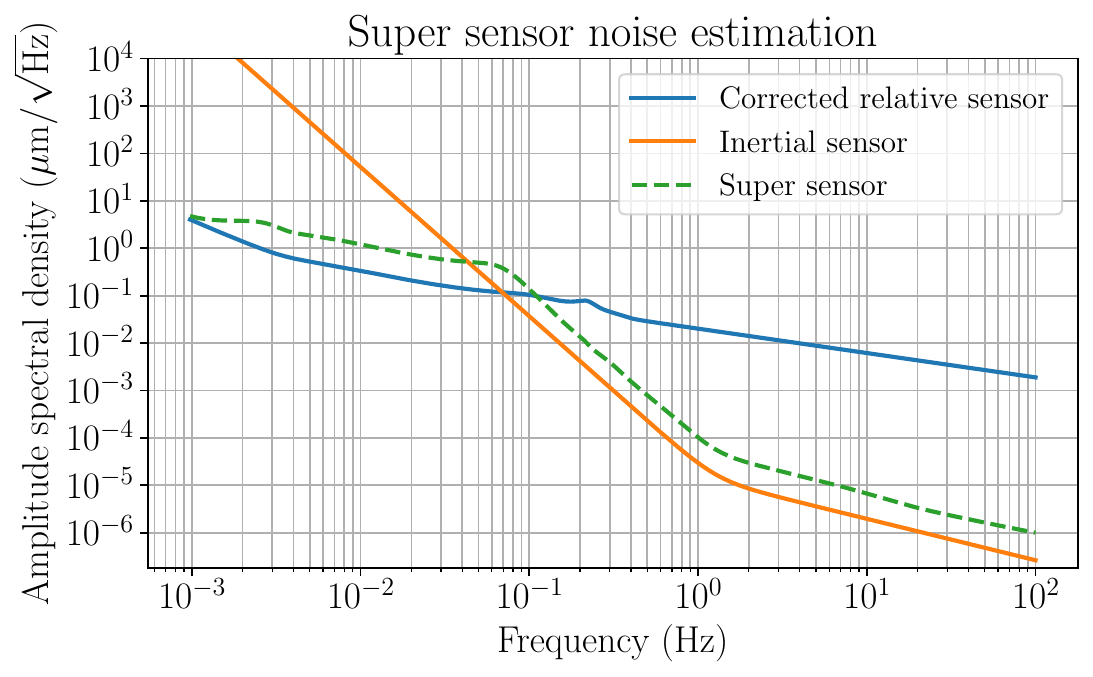}
			\caption{Relative and inertial sensor noises.}
			\label{fig:sensorfusionestimation}
		\end{subfigure}
		\begin{subfigure}{.49\linewidth}
			\centering
			\includegraphics[width=\linewidth]{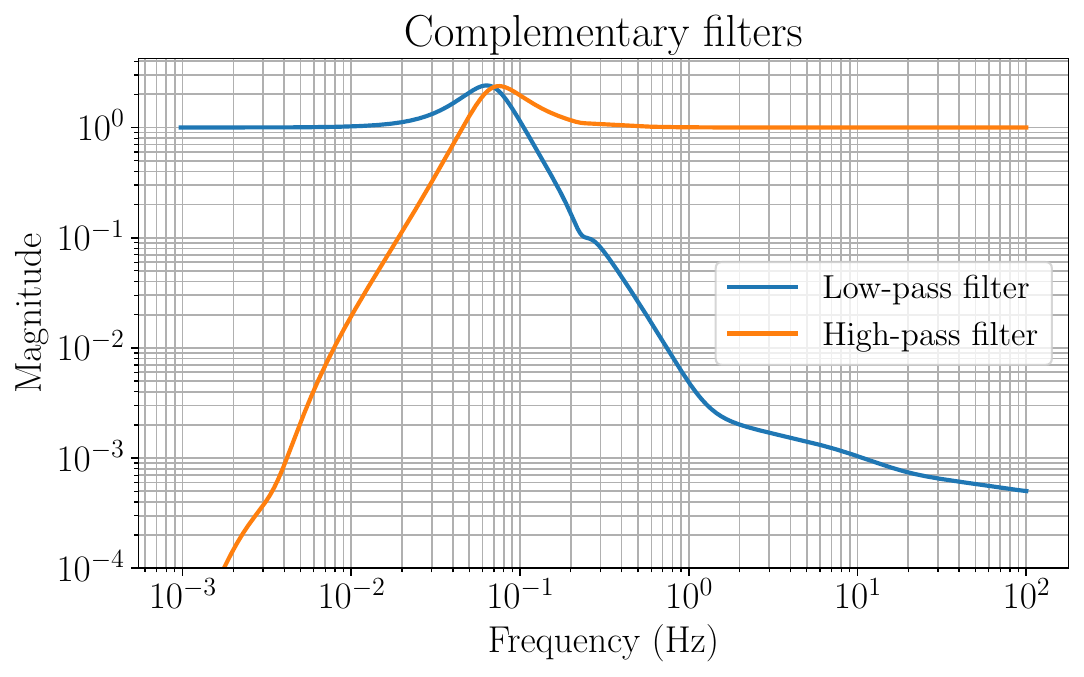}
			\caption{H-infinity optimal complementary fliters.}
			\label{fig:sensorfusionfilter}
		\end{subfigure}
		\caption{Noise spectrums and control filters in the sensor fusion scheme.}
	\end{figure}
	
	Fig.~\ref{fig:sensorfusionestimation} shows the sensor noise of the sensor-corrected relative sensor and the inertial sensor.
	As observed, the inertial sensor exhibits excellent high-frequency noise performance, making it well-suited for active isolation.
	However, it has significantly higher noise than the relative sensor at low frequencies, necessitating high-pass filtering.
	
	Based on these sensor noise characteristics, the complementary filters were optimized using the H-infinity method and are shown in Fig.~\ref{fig:sensorfusionfilter}.
	The crossover of the two filters occurs at approximately 0.07 Hz, aligning with the crossover point of the sensor noise spectra.
	The estimated combined sensor noise is represented by the dashed line in Fig.~\ref{fig:sensorfusionestimation}.
	As observed, the complementary filters are optimally shaped, ensuring that the super sensor noise closely follows the instrumentation limit, with a deviation of $\gamma = 3.43$.
	
	The RMS value of the combined sensor noise is 0.199 $\mu\mathrm{m}$, which is higher than that of the sensor-corrected relative sensor at 0.120 $\mu\mathrm{m}$.
	This increase is due to the reinjection of inertial sensor noise at lower frequencies, a common occurrence in active isolation systems in gravitational-wave detectors.
	
	In terms of the band-limited RMS (BLRMS) between 0.1 and 0.5 Hz, corresponding to the secondary microseism band, the BLRMS value of the combined sensor noise is 0.0188 $\mu\mathrm{m}$.
	For comparison, the BLRMS values of the uncorrected and sensor-corrected relative sensors are 0.159 $\mu\mathrm{m}$ and 0.0367 $\mu\mathrm{m}$, respectively.
	The BLRMS value of the super sensor is approximately 88.2\% and 48.8\% lower than those of the uncorrected and sensor-corrected relative sensor, respectively, indicating greater robustness against secondary microseismic disturbances.
	This makes the super sensor highly suitable for active seismic isolation, even with the increased RMS noise.
	Moreover, its superior noise performance at higher frequencies is particularly beneficial for reducing the overall motion of the suspended optic, especially in mitigating resonances above 0.1 Hz.
	
	\section{Discussions\label{sec:discussions}}
	\subsection{Cascading H-infinity Norms}
	As demonstrated in Sec.~\ref{sec:results}, the H-infinity method has successfully optimized sensor correction and complementary filters individually, pushing performance toward the instrumentation limit.
	While each control configuration individually maintains a constant maximum deviation above the instrumentation limit across all frequencies, this is not necessarily the case for a cascaded system.
	
	In fact, the active isolation system depicted in Fig.~\ref{fig:active_isolation} can be viewed as a cascade of three control systems, each governed by a separate controller: the sensor correction filter, the complementary filters, and the feedback controller.
	The H-infinity optimized sensor correction configuration resulted in a sensor-corrected relative sensor with noise 1.033 times above the instrumentation limit.
	This H-infinity norm carries over to the sensor fusion problem, compounding with the H-infinity norm of the sensor fusion solution, which is $\gamma = 3.43$.
	However, the deviation of 3.43 is relative to the noise of the corrected relative sensor, which is not necessarily the true instrumentation limit of the cascaded system.
	For instance, at certain frequencies, the maximum deviation of the combined sensor noise can reach $1.033 \times 3.43 = 3.54$ times the actual instrumentation limit.
	
	When individually optimizing control filters in a cascaded configuration, the H-infinity norms accumulate, leading to overall performance that is not instrumentation optimal.
	This is also why the optimization of the feedback controller was not demonstrated in Sec.~\ref{sec:results}, as the stacked H-infinity norm would result in a performance too far from the optimal limit.
	To resolve this, the generalized plant must include all control configurations, enabling the co-optimization of all control filters.
	This approach effectively transforms the controller in the generalized plant into a multiple-input, single-output controller, ensuring that the overall performance remains optimal relative to the instrumentation limit.
	
	\subsection{Adaptive Seismic Control}
	The results presented in Sec.~\ref{sec:results} are based on a specific seismic condition.
	In reality, seismic conditions undergo seasonal variations and fluctuate with the weather.
	For instance, the secondary microseismic peak can shift between 0.1 and 0.3 Hz, with its peak magnitude varying from 0.1 to 10 $\mu\mathrm{m}$, spanning two orders of magnitude.
	Since the H-infinity method relies on an optimal trade-off between noise injection and seismic noise suppression, control filters optimized for a particular seismic condition may not remain optimal under different conditions.
	
	Elevated secondary microseismic disturbances and distant earthquakes remain some of the most challenging seismic conditions for gravitational-wave detectors.
	Previous studies have shown that real-time adjustments to the sensor correction filter can effectively mitigate the effects of distant earthquakes, helping to maintain interferometer alignment during abrupt disturbances \cite{Schwartz_2020}.
	This concept can be extended to a more general adaptive control approach, where control filters are dynamically selected based on real-time seismic conditions to address varying seismic noise.
	In this framework, the H-infinity method serves as a crucial tool for designing a pool of control filters tailored to different seismic scenarios.
	This pool then forms a parameter space in which real-time optimization occurs, enabling adaptive control.
	
	\section{Conclusions\label{sec:conclusions}}
	
	This work applied the H-infinity method to optimize sensor correction and complementary filters for active seismic isolation in a gravitational-wave detector.
	By formulating sensor correction, sensor fusion, and feedback control within a unified optimization framework, we demonstrated that each control scheme effectively minimized sensor noise while maintaining performance near the instrumentation limit.
	The successful implementation of these methods in KAGRA led to significant improvements in seismic isolation, contributing to an increase in the detector's duty cycle from 53\% in O3GK to 80\% in O4a.
	
	However, when these control schemes are applied in a cascaded configuration, their individual H-infinity norms compound, leading to a deviation from the true optimal performance.
	To address this, we proposed a co-optimization approach that integrates all control configurations into a single generalized plant, ensuring that the overall system remains optimal relative to the instrumentation limit.
	
	Additionally, we discussed the impact of varying seismic conditions on control performance and highlighted the need for adaptive control strategies.
	By leveraging the H-infinity method to design a pool of control filters for different seismic environments, real-time optimization can be achieved, allowing the system to dynamically adjust to changing seismic conditions.
	Future work will focus on implementing and testing adaptive control strategies in real-time, ensuring optimal performance under dynamic seismic disturbances and improving the robustness of gravitational-wave detectors.
	
	\ack
		This material is based upon work supported by NSF’s LIGO Laboratory which is a major facility fully funded by the National Science Foundation.
		
		This work was supported by Ministry of Education, Culture, Sports, Science and Technology (MEXT), Japan Society for the Promotion of Science (JSPS) in Japan; National Research Foundation (NRF) and Ministry of Science and ICT (MSIT) in Korea; Academia Sinica (AS) and National Science and Technology Council (NSTC) in Taiwan.
		
		This work was supported by the Science and Technology Facilities Council grant number ST/V005618/1.

	\section*{References}
	\bibliographystyle{unsrt}
	\bibliography{bibliography.bib}
\end{document}